%% file: main.tex
\documentclass[10pt,conference]{IEEEtran}
\IEEEoverridecommandlockouts
\input{def_miba}

\def\BibTeX{{\rm B\kern-.05em{\sc i\kern-.025em b}\kern-.08em
    T\kern-.1667em\lower.7ex\hbox{E}\kern-.125emX}}

\begin{document}

\title{Model Order Selection with Variational Autoencoding
\thanks{This work is funded by the Bavarian Ministry of Economic Affairs, Regional Development and Energy within the project 6G Future Lab Bavaria. The authors acknowledge the financial support by the Federal Ministry of Education and Research of Germany in the programme of ``Souverän. Digital. Vernetzt.''. Joint project 6G-life, project identification number: 16KISK002}
}

\author{\IEEEauthorblockN{Michael Baur, Franz Weißer, Benedikt Böck, and Wolfgang Utschick}
\IEEEauthorblockA{TUM School of Computation, Information and Technology, Technical University of Munich, Germany}
Email: \{mi.baur, franz.weisser, benedikt.boeck, utschick\}@tum.de
}
	\maketitle
    
    \thispagestyle{cfooter}
    
	\begin{abstract}
        Classical methods for model order selection often fail in scenarios with low SNR or few snapshots. Deep learning-based methods are promising alternatives for such challenging situations as they compensate lack of information in the available observations with training on large datasets. This manuscript proposes an approach that uses a variational autoencoder (VAE) for model order selection. The idea is to learn a parameterized conditional covariance matrix at the VAE decoder that approximates the true signal covariance matrix. The method is unsupervised and only requires a small representative dataset for calibration after training the VAE. Numerical simulations show that the proposed method outperforms classical methods and even reaches or beats a supervised approach depending on the considered snapshots.
	\end{abstract}

\input{introduction}
	\input{system}
	\input{mo}
	\input{results}
	\input{conclusion}
	
	\bibliographystyle{IEEEtran}
	\bibliography{main}
	
\end{document}

%% file: def_miba.tex
\usepackage{tikz}
\usepackage{pgfplots}
\usepackage{bm}
\usepackage{amsmath} 
\usepackage{amsfonts}
\usepackage{graphicx}
\usepackage{siunitx}
\usepackage{color}
\usepackage{fancyhdr}
\usepackage{array,multirow}
\usepackage{adjustbox}
\usepackage{balance}
\usepackage{cite}
\usepackage[acronym,shortcuts]{glossaries}
\usetikzlibrary{calc, shapes.geometric}

\ifCLASSOPTIONcompsoc
\usepackage[caption=false,font=normalsize,labelfont=sf,textfont=sf]{subfig}
\else
\usepackage[caption=false,font=footnotesize]{subfig}
\fi

\pgfplotsset{
	discard if not/.style 2 args={
		x filter/.code={
			\edef\tempa{\thisrow{#1}}
			\edef\tempb{#2}
			\ifx\tempa\tempb
			\else
			\fi
		}
	}
}

\fancypagestyle{cfooter}{ %
\fancyhf{} 
\cfoot{\scriptsize{© 2023 IEEE. Personal use of this material is permitted. Permission from IEEE must be obtained for all other uses, in any current or future media, including reprinting/republishing this material for advertising or promotional purposes, creating new collective works, for resale or redistribution to servers or lists, or reuse of any copyrighted component of this work in other works.}}

}

\pgfplotsset{compat=1.15}
\definecolor{mylila}{RGB}{153,50,204} 
\definecolor{mygreen}{RGB}{176,191,26}

\newcommand{\vmu}{\bm{\mu}}
\newcommand{\vsig}{\bm{\sigma}}

\newcommand{\vaec}{{VAE-$\vc$}}
\newcommand{\vaee}{{VAE-$\ve$}}
\newcommand{\vaecsig}{{VAE-$\vc$-$\sigma^2$}}
\newcommand{\vaeesig}{{VAE-$\ve$-$\sigma^2$}}

\newcommand{\va}{{\bm{a}}}

\newcommand{\vc}{{\bm{c}}}

\newcommand{\ve}{{\bm{e}}}

\newcommand{\vn}{{\bm{n}}}

\newcommand{\vs}{{\bm{s}}}

\newcommand{\vy}{{\bm{y}}}
\newcommand{\vz}{{\bm{z}}}

\newcommand{\ma}{{\bm{A}}}

\newcommand{\mc}{{\bm{C}}}

\newcommand{\mf}{{\bm{F}}}

\newcommand{\mr}{{\bm{R}}}

\newcommand{\CC}{\mathbb{C}}

\newcommand{\jim}{\mathrm{j}}

\DeclareMathOperator{\E}{E}
\DeclareMathOperator{\diag}{diag}
\DeclareMathOperator{\KL}{D_{KL}}
\DeclareMathOperator{\tr}{tr}
\DeclareMathOperator*{\tran}{{^{\mkern-2mu{T}}}}
\DeclareMathOperator*{\herm}{^{\mkern-2mu{H}}}

\newacronym{tdd}{TDD}{time division duplex}
\newacronym{fdd}{FDD}{frequency division duplex}
\newacronym{lmmse}{LMMSE}{linear minimum mean squared error}
\newacronym{mse}{MSE}{mean squared error}
\newacronym{mmse}{MMSE}{minimum mean squared error}
\newacronym{nmse}{NMSE}{normalized mean squared error}
\newacronym{mimo}{MIMO}{multiple-input multiple-output}
\newacronym{simo}{SIMO}{single-input multiple-output}
\newacronym{miso}{MISO}{multiple-input single-output}
\newacronym{siso}{SISO}{single-input single-output}
\newacronym{deep}{DL}{deep learning}
\newacronym{ofdm}{OFDM}{orthogonal frequency division multiplexing}
\newacronym{csi}{CSI}{channel state information}
\newacronym{ula}{ULA}{uniform linear array}
\newacronym{ura}{URA}{uniform rectangular array}
\newacronym{dft}{DFT}{discrete fourier transform}
\newacronym{bs}{BS}{base station}
\newacronym{mt}{MT}{mobile terminal}
\newacronym{ae}{AE}{autoencoder}
\newacronym{ml}{ML}{machine learning}
\newacronym{doa}{DoA}{direction of arrival}
\newacronym{dod}{DoD}{direction of departure}
\newacronym{kl}{KL}{Kullback-Leibler}
\newacronym{iid}{i.i.d.}{independent and identically distributed}
\newacronym{fc}{FC}{fully connected}
\newacronym{nn}{NN}{neural network}
\newacronym{cnn}{CNN}{convolutional neural network}
\newacronym{ls}{LS}{least squares}
\newacronym{snr}{SNR}{signal-to-noise ratio}
\newacronym{ce}{CE}{channel estimation}
\newacronym{ul}{UL}{uplink}
\newacronym{mpc}{MPCs}{multipath components}
\newacronym{rt}{RT}{ray-tracing}
\newacronym{mmd}{MMD}{maximum mean discrepancy}
\newacronym{cdf}{CDF}{cumulative distribution function}
\newacronym{tpr}{TPR}{true positive rate}
\newacronym{dl}{DL}{deep learning}
\newacronym{mo}{MO}{model order}
\newacronym{ic}{IC}{information criteria}
\newacronym{vae}{VAE}{variational autoencoder}
\newacronym{gmm}{GMM}{Gaussian mixture model}
\newacronym{elbo}{ELBO}{evidence lower bound}
\newacronym{aic}{AIC}{Akaike information criterion}
\newacronym{bic}{BIC}{Bayesian information criterion}
\newacronym{mdl}{MDL}{maximum description length}
\newacronym{cc}{CC}{convolutional channel}
\newacronym{cl}{CL}{convolutional layer}
\newacronym{ll}{LL}{linear layer}
\newacronym{rl}{RL}{reshaping layer}
\newacronym{bn}{BN}{batch normalization}

\newacronym{kde}{KDE}{kernel density estimate}

%% file: introduction.tex
\section{Introduction}
\label{sec:intro}

\Ac{mo} selection determines the number of impinging wavefronts incident at a receiver.
The \ac{mo} is an essential quantity for \ac{doa} estimation, both for classical~\cite{Krim1996} and current \ac{dl} methods~\cite{Barthelme2021}.
Most well-known \ac{mo} selection approaches utilize \ac{ic}~\cite{Stoica2004}. 
More current treatment of model selection is covered in~\cite{Ding2018}, where \ac{dl} methods are left out, however. 

A popular \ac{ic}-based method for \ac{mo} selection reaches back to the 80s~\cite{Wax1985}. 
The method is based on a subspace decomposition of the sample covariance matrix and performs well in cases with high \ac{snr} and many snapshots. 
In contrast, for low \ac{snr} or few snapshots, the sample covariance matrix is a bad estimate of the true covariance matrix. 
Consequently, the method fails in these cases. 
\Ac{dl} methods are promising candidates to perform well in such difficult situations. 
As a result of the repeated presentation of samples during the offline training phase, the \ac{dl} model extracts overall prior information of the data and can compensate for lack of knowledge in observations during the deployment phase, e.g., if only a few snapshots are available. 
The strong performance of \ac{dl}-based methods is demonstrated in~\cite{Bialer2019,Barthelme2020,Barthelme2021a,Zhou2022}, which use relatively simple neural network architectures to determine the \ac{mo} based on the (preprocessed) snapshots. 
The methods are supervised, requiring access to a dataset, where observations are labeled with their corresponding \ac{mo}. 

If the exact signal model and the exact model of the antenna array were available, it would be possible to generate unlimited amounts of labeled data. 
This assumption, however, only holds for idealistic circumstances, e.g., calibrated antenna arrays, that do not hold in reality. 
Under realistic conditions, data would only be available in the form of measurements without any labels.
These aspects motivate the investigation of unsupervised learning methods because they do not require labeling.
Unsupervised methods additionally offer to include model imperfections in the framework directly.
The \ac{vae} is an unsupervised framework that learns the data distribution by maximizing a lower bound to the data log-likelihood~\cite{Kingma2014}. 
It belongs to the class of generative models, which means that the model can generate entirely new samples from the learned distribution. 
Despite its popularity in image processing and related disciplines, the \ac{vae} is rarely employed in communications tasks. 
A current publication investigates the generative modeling performance of the \ac{vae} in a millimeter-wave UAV scenario~\cite{Xia2022a}. 
Channel equalization is another domain where the \ac{vae} is applied successfully~\cite{Caciularu2018,Caciularu2020,Lauinger2022a}, as well as channel estimation~\cite{Baur2022}. 

Motivated by the performance of the \ac{vae} channel estimator in~\cite{Baur2022}, we propose a method for \ac{mo} selection based on a \ac{vae}. 
Our method is unsupervised, and only a small representative dataset is required to distinctly assign the \ac{mo} after training of the \ac{vae}. 
The method is supposed to fill the gaps where classical methods for \ac{mo} selection fail, i.e., at low \ac{snr} and few snapshots. 
The contributions are as follows.
By parameterizing the covariance matrix of the conditionally Gaussian distribution at the \ac{vae} decoder with an oversampled \ac{dft} matrix, we can learn an approximation of the eigenvalues of the true signal covariance matrix. 
We leverage the approximation to determine the \ac{mo} with a custom evaluation routine based on entropy. 
Numerical simulations show the advantage of the proposed framework over \ac{ic}-based methods in the considered scenarios. 
Moreover, the proposed method can beat a supervised \ac{mo} selection method in a single snapshot scenario.

%% file: system.tex
\section{System Model}
\label{sec:system}

An antenna array with $N$ elements receives signals from $L$ sources in the far field, which characterize the impinging wavefronts. 
The received signal vector $\vy(t)\in\CC^N$ at snapshot $t$ of in total $T$ snapshots is expressed as
\begin{equation}
	\vy(t) = \ma(\bm{\theta}) \vs(t) + \vn(t), \quad t=1,\ldots,T
	\label{eq:system}
\end{equation}
with the array manifold $ \ma(\bm{\theta})\in\mathbb{C}^{N\times L}$, the DoA $\bm{\theta}$, the transmitted signal $\vs(t)\sim\mathcal{N}_\mathbb{C}(\bm{0},\mc_{\vs})$, and additive white Gaussian noise $\vn(t)\sim\mathcal{N}_\mathbb{C}(\bm{0},\sigma_n^2\mathbf{I})$. 
The columns of the array manifold are defined by the steering vectors of the employed array geometry and evaluated at the respective angles. 
Transmitted signals are assumed to originate from uncorrelated sources with different powers, i.e., the matrix $\mc_{\vs}$ is diagonal with positive and potentially dissimilar elements on its diagonal. 
The covariance matrix of the received signal is 
\begin{equation}
    \mc_{\vy}=\mr+\sigma_n^2\mathbf{I}, \qquad\mr=\ma(\bm{\theta})\mc_{\vs}\ma(\bm{\theta})\herm.
\end{equation}
Note that this result only holds for a fixed $\bm{\theta}$, which is assumed to be constant over the snapshots.

We consider a \ac{ula} with half-wavelength spacing. The steering vector of a \ac{ula} at angle~$\theta$ is $\va(\theta)=[1, \exp(\jim \pi \sin(\theta)), \ldots, \exp(\jim \pi (N-1) \sin(\theta))]\herm$.
We furthermore set $\tr(\mc_{\vs})=1$, which allows us to define the \ac{snr} as $1/\sigma_n^2$.

%% file: mo.tex
\section{Model Order Selection Techniques}
\label{sec:mo}

The goal of \ac{mo} selection in a typical \ac{doa} scenario is to determine the number of sources $L$. 
In principle, the number of non-zero eigenvalues of $\mr$ gives the \ac{mo}. 
It is evident that $\mr$ is not accessible during operation, which requires us to leverage the available snapshots $\vy(t)$ to determine the \ac{mo}, as it is the only information we receive, besides structural information, e.g., of the antenna array.

\subsection{Information Criteria}
\label{subsec:ic}

Classical approaches for \ac{mo} selection are often based on \ac{ic}~\cite{Stoica2004}. 
They evaluate the quality of a model for given data and account for the degrees of freedom by an additive penalty term. 
Among the most well-known \ac{ic} rules are the \ac{aic} or the \ac{bic}.
The latter is also known as \ac{mdl}, the term we use in this work.

An approach to apply the \ac{aic} and \ac{mdl} to the \ac{mo} selection task for the system model in~\eqref{eq:system} is presented in~\cite{Wax1985}. 
The authors compute a maximum likelihood estimate for every \ac{mo}. Afterward, they obtain the \ac{mo} by a subspace decomposition of the sample covariance matrix  $\hat{\mc} = \frac{1}{T}\sum_{t=1}^{T} \vy(t)\vy(t)\herm.$
The performance of such an approach is highly dependent on the quality of $\hat{\mc}$ in estimating the true covariance matrix of the snapshots. 
Consequently, for a low number of snapshots or low \ac{snr}, the \ac{aic} and \ac{mdl} methods, as they are defined in~\cite{Wax1985}, are error-prone in determining the correct \ac{mo}. 
For cases where $\hat{\mc}$ is of good quality, the \ac{ic} based methods perform well. 
Therefore, this work aims to develop a method for \ac{mo} selection that works well in cases not covered by \ac{ic} and is additionally unsupervised.

\subsection{Variational Autoencoder Preliminaries}
\label{subsec:vae}

\begin{figure}[t]
	\centering
	\includegraphics[width=\linewidth]{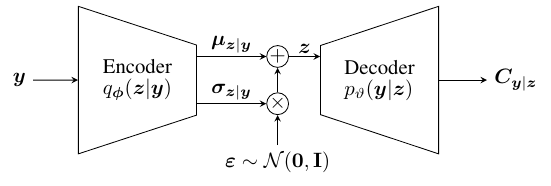}
	\vspace*{-12pt}
	\caption{Sketch of a \ac{vae} that only learns a covariance matrix $\mc_{\vy|\vz}$ at the decoder for an input vector $\vy$ at the encoder. The encoder and decoder are each realized as neural networks, parameterized by their model weights $\phi$ and $\vartheta$, respectively.}
	\label{fig:vae}
	\vspace*{-6pt}
\end{figure}

Consider a \ac{vae}~\cite{Kingma2014} as Fig.~\ref{fig:vae} shows it. 
A data sample~$\vy$ is put into the encoder to yield a sample $\vz$ of the variational distribution $q_{\phi}(\vz|\vy)$, referred to as a latent sample. 
The latent sample is fed into the decoder to obtain a conditional covariance matrix $\mc_{\vy|\vz}$. 
The training objective of a \ac{vae} during stochastic optimization is to maximize the \ac{elbo}, which is given by
\begin{equation}
	\mathcal{L}_{\phi,\vartheta}(\vy) = \E_{q_{\phi}(\vz|\vy)}\left[\log p_{\vartheta}(\vy|\vz)\right] - \KL(q_{\phi}(\vz|\vy)\,\|\,p(\vz)).
	\label{eq:vae}
\end{equation}
The \ac{elbo} is a lower bound to the data log-likelihood.
The outcome of the \ac{elbo} maximization are estimates for the distributions $q_{\phi}(\vz|\vy)$ and $p_{\vartheta}(\vy|\vz)$. 
A common choice for the involved probability distributions in the \ac{elbo} is Gaussians. 
Thus, we set $p(\vz)=\mathcal{N}(\bm{0},\mathbf{I})$, $p_{\vartheta}(\vy|\vz)= \mathcal{N}_{\CC}(\bm{0},\mc_{\vy|\vz})$, and $q_{\phi}(\vz|\vy)=\mathcal{N}(\vmu_{\vz|\vy},\diag(\vsig^2_{\vz|\vy}))$. 
We can compute the \ac{elbo} in closed-form with these definitions, which is explained in detail in~\cite{Baur2022}. 
As a result, we have $\mathcal{L}_{\phi,\vartheta}(\vy) =$
\begin{equation}
	\frac{1}{2}\bm{1}\tran\left( \log\vsig^2_{\vz|\vy} - \vmu^2_{\vz|\vy} - \vsig^2_{\vz|\vy} \right) -\log\det\mc_{\vy|\vz} - \vy\herm\mc_{\vy|\vz}^{-1}\vy
\end{equation}
as optimization objective, with the all-ones vector $\bm{1}$. 
We obtain $\vmu_{\vz|\vy}$ and $\vsig_{\vz|\vy}$ with the encoder and $\mc_{\vy|\vz}$ with the decoder.
The subscripts $\phi$ and $\vartheta$ correspond to the encoder and decoder neural network weights, respectively, and parameterize $\vmu_{\vz|\vy}$, $\vsig_{\vz|\vy}$, and $\mc_{\vy|\vz}$.
Finally, using the reparameterization trick to obtain the latent vector $\vz$ yields the \ac{vae} proposed in the related literature~\cite{Kingma2014}.  
For further background details, we refer the reader to~\cite{Baur2022,Kingma2019}. 
In the former~\cite{Baur2022}, the authors describe a similar application case.

\subsection{Model Order Selection by Variational Autoencoding}
\label{subsec:mo-vae}

Recall that the \ac{vae} maximizes a lower bound to the data log-likelihood. 
An intuitive way to determine the \ac{mo} in combination with a \ac{vae} is to work with the conditional covariance matrix at the decoder $\mc_{\vy|\vz}$. 
If it is a good estimate of the actual covariance matrix $\mc_{\vy}$, it should be possible to infer the \ac{mo} based on $\mc_{\vy|\vz}$. 
Our choice for the distribution $p_{\vartheta}(\vy|\vz)= \mathcal{N}_{\CC}(\bm{0},\mc_{\vy|\vz})$ differs from the standard version in the \ac{vae} literature. 
The conditional covariance matrix is usually either assumed to be diagonal or a scaled identity. 
Furthermore, the mean value of $p_{\vartheta}(\vy|\vz)$ is commonly a learnable non-zero vector.
While this choice achieves good results, e.g., for generating new samples, it is unsuited for our purposes because the true covariance matrix has a rich structure due to the given antenna array. 
For instance, the covariance matrix is Toeplitz structured for a \ac{ula} and uncorrelated sources, which we cannot model with the standard version of the \ac{vae}.

To this end, the authors in~\cite{Baur2022} parameterize the channel covariance matrix as a circulant matrix $\mf\herm\diag(\tilde\vc)\mf$ with a \ac{dft} matrix $\mf\in\mathbb{C}^{N\times N}$ and $\tilde\vc\in\mathbb{R}_+^N$ to approximate the Toeplitz channel covariance matrix of a \ac{ula}. 
The parameterization performs very well for channel estimation. 
However, in our experiments for \ac{mo} selection, we discovered that the eigenvalues of $\mc_{\vy|\vz}$, which are $\tilde\vc+\sigma_n^2\bm{1}$ with the circulant parameterization, are not representative for the \ac{mo}. 
Unfortunately, this property prevents the direct application of \ac{ic} rules to determine the \ac{mo}. 
For example, a sample with a true \ac{mo} of one might have a single eigenvalue larger than the noise variance and other smaller but non-negligible eigenvalues in addition to the single dominant eigenvalue. 
This relation comes from the fact that the row vectors of $\mf$ are steering vectors of the \ac{ula} evaluated at different angles. 
Hence, the learned eigenvalues express the variance of the signal space in terms of the available steering vectors in $\mf$. 
If an eigendirection is not collinear to one of the steering vectors, it is represented as a linear combination of the Fourier basis with more basis vectors than the actual \ac{mo} tells.
Consequently, we have more eigenvalues greater than the noise variance as the actual \ac{mo}.

One possible adaption to the circulant parameterization to cope with this problem is to use an oversampled \ac{dft} matrix $\tilde{\mf}\in\mathbb{C}^{KN\times N}$ with an oversampling factor $K\in\mathbb{N}_+$. 
Such a matrix contains a finer grid of directions in the signal space, which allows us to put more energy into fewer eigenvalues. 
The drawback is that the right-inverse of $\tilde{\mf}$ does not exist. 
Only the left-inverse exists, which is $\tilde{\mf}\herm$. 
The non-existence of the right-inverse results in higher computational complexity during the training phase because the inverse of the conditional covariance matrix
\begin{equation}
	\mc_{\vy|\vz}=\tilde{\mf}\herm\diag(\vc_{\vy|\vz})\tilde{\mf} = \tilde{\mf}\herm\diag(\vc + \sigma_n^2\bm{1})\tilde{\mf},
	\label{eq:cov-dec}
\end{equation}
with $\vc_{\vy|\vz},\vc \in \mathbb{R}_+^{KN}$, must be computed for every $\vy$. 
Note that $\vc$ is the output of the decoder.
However, this is acceptable as all additional effort happens during the training phase and can be done offline and in advance. 
So, since the \ac{ic} rules are not directly applicable, the question arises: how can we determine the \ac{mo} with the described \ac{vae} that we train on unlabeled data originating from the system model in~\eqref{eq:system}?

Since the eigenvalues of $\tilde{\mf}\herm\diag(\vc)\tilde{\mf}$ are representative for the \ac{mo} and defined by the vector $\vc$, which determines the weighting of the rows in $\tilde\mf$, it should be possible to determine the \ac{mo} based on $\vc$. 
However, since $\tilde{\mf}$ is an oversampled \ac{dft} matrix, it is still not possible to say that, e.g., $m$ dominant values in $\vc$ allow to infer an \ac{mo} of $m$. 
Instead, the energy distribution over $\vc$ is tied to the \ac{mo}. 
We should therefore interpret $\vc$ as a feature vector. 
Normalizing the vector by its sum yields a non-negative vector that sums to one. 
However, a criterion to separate the normalized vectors based on their \ac{mo} is needed. 
The entropy is a suitable candidate for this task as it measures how evenly the energy is distributed over the values in $\vc$. 
The entropy is high if the total energy is distributed over many values of $\vc$, which is a sign of a high \ac{mo}. 
In contrast, the entropy is small if almost all the energy concentrates on one value, indicating a low \ac{mo}. 

\begin{figure}[t]
	\centering
	\includegraphics[width=\linewidth]{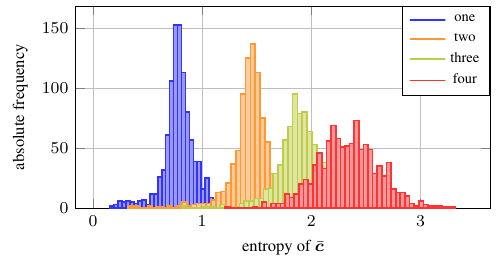}
	\vspace*{-20pt}
	\caption{Histogram of the entropies of $\bar\vc$ at an \ac{snr} of 10\,dB for \ac{mo} one to four. In this figure, \textit{five} snapshots are considered.}
	\label{fig:entropy}
	\vspace*{-10pt}
\end{figure}

In our experiments, it was better to learn a covariance matrix at the decoder for every snapshot separately and then average the $\vc$ after the training than to learn $\vc$ for all snapshots jointly. 
More precisely, let the vector $\vc(t)$ be obtained at the VAE decoder for the encoder input $\vy(t)$.
In the next step, the quantities 
\begin{equation}
    \hat{\vc} = \frac{1}{T} \sum_{t=1}^T \vc(t), \quad \bar\vc = \frac{\hat{\vc}}{\sum_{i=1}^{KN}\hat{c}_{i}}
    \label{eq:cbar}
\end{equation}
are computed, which averages the decoder outputs over the snapshots and normalizes the resulting vector such that the sum of the elements is one.
Afterward, we calculate the entropy of $\bar\vc$ by treating all elements $\bar{c}_{i}$ in the vector as probabilities for different outcomes of a random variable, i.e.,
\begin{equation}
    \mathrm{H}(\bar\vc) = \sum_{i=1}^{KN} - \bar{c}_{i} \log( \bar{c}_{i} ).
    \label{eq:entropy}
\end{equation}
If we calculate the entropies according to this procedure, we obtain a histogram as in Fig.~\ref{fig:entropy}. 
An ex-post view shows that, from left to right, the different colors belong to \ac{mo} one to four. 
We excluded the data of \ac{mo} zero because we find it by comparing the highest value in $\hat\vc$ with the noise variance, which does not require further steps. 
Consequently, a value smaller than $\sigma_n^2$ refers to \ac{mo} zero. 
The histogram shows that entropies can provide a good measure of the classification of observations in terms of their underlying \ac{mo}. 
What remains is to find suitable thresholds for entropy values that allow the determination of the \ac{mo}. 
More precisely, since we want to remain unsupervised, we use a one-dimensional \ac{gmm} to model the entropy distribution to obtain thresholds.
Finally, we determine the \ac{mo} with the \ac{gmm} component corresponding to the calculated entropy value.
Exemplarily, in Fig.~\ref{fig:entropy}, an entropy value of $1.4$ would be assigned to the \ac{gmm} component with the second largest mean value, which represents \ac{mo} two.
The proposed method is not limited to using a \ac{gmm}.
Any tool that yields thresholds for entropy values to assign the correct \ac{mo} may be used.

%% file: results.tex
\section{Results}
\label{sec:results}

This section first provides details regarding implementing the \ac{vae} architecture. 
Afterward, we present \ac{mo} selection results for the proposed method for one and five snapshots.

\subsection{Implementation Details}
\label{subsec:impl}

The employed architectures follow the principles in~\cite{Baur2022}. 
Since the snapshots are complex-valued, we stack their real and imaginary parts to create input vectors with two \acp{cc}. 
We decided to incorporate \acp{cl} in the neural networks inside the \ac{vae} as they have demonstrated superior performance compared to networks solely built from \acp{ll} in vast amounts of prior work.
The neural networks in every \ac{vae} have the following architectures. 
The encoder consists of three times a building block of a \ac{cl}, a \ac{bn} layer, and a ReLU activation function. 
Each \ac{cl} has kernel size seven. 
The input samples are mapped from two to 64, to 128, and to 192 \acp{cc}. 
We use a stride of one.
The three blocks are followed by two \acp{ll} that map to $\vmu_{\vz|\vy}$ and $\vsig_{\vz|\vy}$, which are of dimensionality 16. 
With $\vmu_{\vz|\vy}$ and $\vsig_{\vz|\vy}$, the reparameterization trick is performed to obtain one sample $\vz$ of $q_{\phi}(\vz|\vy)$.
The decoder architecture is analog to the encoder architecture, just flipped symmetrically and transposed \acp{cl} are used instead of standard \acp{cl}. 
At the output of the decoder, a \ac{ll} maps to $\vc$.
We found the architecture with a random search over the hyperparameter space for the configuration that yields the highest \ac{elbo} value with the \textit{Tune} package~\cite{Liaw2018}. 
We implement our neural networks with the help of \textit{PyTorch} and optimize them with the \textit{Adam} optimizer~\cite{Kingma2015} with a learning rate of $10^{-4}$. 
The \acp{vae} are trained for an \ac{snr} range from -16 to 26\,dB until the \ac{elbo} saturates to also include border \ac{snr} samples. 
After the training, we use the model that yields the highest \ac{elbo} value. 
Note that the training with an \ac{snr} range implies that the proposed approach is \ac{snr} independent.

\subsection{Numerical Simulations}
\label{subsec:sim}

\begin{figure}[t]
	\includegraphics{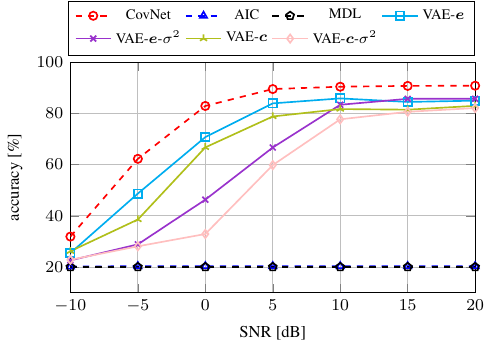}
	\caption{Numerical results for \ac{mo} selection with \textit{five} snapshots. The proposed methods are displayed with solid lines.}
	\label{fig:5-obs}
\end{figure}

\begin{figure}[t]
	\includegraphics{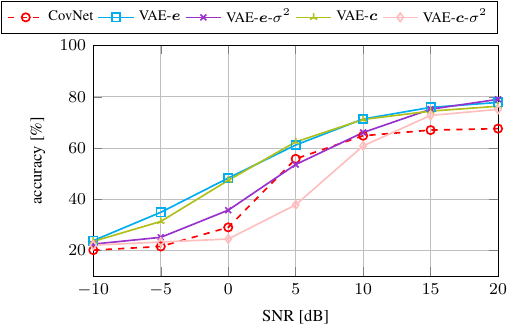}
	\caption{Numerical results for \ac{mo} selection with \textit{one} snapshot. The proposed methods are displayed with solid lines.}
	\label{fig:1-obs}
\end{figure}

We create training and evaluation data according to the system model in~\eqref{eq:system} for \ac{mo} zero to four.
During the training phase, we create $10^4$ new samples per \ac{mo} in every epoch and train with a batch size of $16$.
Each training sample consists of ten snapshots. 
After training, we create another $10^3$ entirely new samples per \ac{mo} for evaluation purposes.
Please note that the numbers $10^4$ and $10^3$ do not refer to the number of snapshots but to the number of samples in the training and evaluation data, respectively.
Furthermore, labeling the samples with their true \ac{mo} relates to the evaluation phase of the proposed methods. 
\ac{vae} training remains unsupervised. 
The information in the plots applies to the number of snapshots in the evaluation samples, i.e., either \textit{five} or \textit{one} snapshots are considered.
Although the number of snapshots in the evaluation data differs from the training data, which is ten, we achieved better results when training with data featuring ten snapshots.
The same \acp{vae} are used for the evaluations in Fig.~\ref{fig:5-obs} and~\ref{fig:1-obs}. 
The entries of $\bm{C}_{\vs}=\diag(\vc_{\vs})$ are sampled from a uniform distribution between $1/8$ and $1$ and afterward normalized such that they sum to one. 
The DoA of the \ac{ula} is sampled from a uniform distribution between $-\pi/2$ and $\pi/2$. 
The oversampling factor is $K=4$, and the number of antennas is $N=64$.

In total, we evaluate four different \ac{vae}-based approaches for \ac{mo} selection. 
The first model employs the method based on the entropy of $\bar\vc$ from~\eqref{eq:cbar} described in Section~\ref{subsec:mo-vae}. 
It assumes knowledge of the noise variance $\sigma_n^2$ and is called \vaec. 
Moreover, to explore the performance gap if working with $\hat\vc$ instead of the eigenvalues of $\tilde{\mf}\herm\diag(\hat\vc)\tilde{\mf}$, we apply the same entropy-based method to the eigenvalues. 
In particular, we apply the method to the eigenvalues of 
\begin{equation}
    \tilde{\mf}\herm\diag(\hat\vc)\tilde{\mf} = \frac{1}{T}\sum_{t=1}^T \tilde{\mf}\herm\diag(\vc(t))\tilde{\mf}.
    \label{eq:avgevd}
\end{equation}
This approach is named \vaee. 
Both \vaec\ and \vaee\ can be implemented as versions where the noise variance is left as an optimization variable. 
We also implement these versions and term them \vaecsig\ and \vaeesig. 
Please note again that, for all of our \ac{vae}-based approaches, the same VAE model is used for every \ac{snr}. 

The proposed method discriminates between different \acp{mo}, but with a possible ambiguity that arises from strongly underrepresented \ac{mo} data.
The ambiguity can be reasoned based on Fig.~\ref{fig:entropy}, where underrepresented \acp{mo} lead to shifts of the entropy thresholds.
Thus, an additional representative dataset with few samples that contains every \ac{mo} remains essential to determine the entropy thresholds of a \ac{mo} after \ac{vae} training.
As outlined in Section~\ref{subsec:mo-vae}, we find the thresholds with a \ac{gmm}.
For our simulations, we assume the evaluation data is representative and can be used for this task. 
During the evaluation of the \acp{vae}, every snapshot $\vy(t)$ is fed into the encoder separately to obtain $\vmu_{\vz|\vy(t)}$, which is put into the decoder to receive the respective $\vc(t)$. 
Hence, we discard $\vsig_{\vz|\vy(t)}$ and skip the sampling process in the latent space. 

We also implement the supervised approach from~\cite{Barthelme2020}, named CovNet, to find out how well our methods perform compared to a supervised method. 
In our experiments, CovNet delivers better evaluation performance on one- and five-snapshot data when we train it on ten-snapshot data, which is why we also train CovNet solely on ten-snapshot data and use this model for all evaluations.

In Fig.~\ref{fig:5-obs}, we display the performance concerning \ac{mo} selection for evaluation data with \textit{five} snapshots. 
The ordinate displays the percentage of correctly determined \ac{mo}s between zero and four, which means that we have five \ac{mo}s in total. 
As can be seen, the \ac{ic} methods A\ac{ic} and \ac{mdl} deteriorate to guessing the \ac{mo} since the sample covariance matrix is of bad quality. 
Furthermore, we observe that the supervised CovNet method outperforms all other methods. 
The \ac{vae}-based methods manage to shrink the performance gap between CovNet at high \ac{snr} but stay a few percentages below CovNet. 
When working with $\hat\vc$ from~\eqref{eq:cbar} instead of the eigenvalues from~\eqref{eq:avgevd}, the curves of \vaec\ and \vaee\ show that a few percentages of accuracy are lost with \vaec. 
The same holds for \vaeesig\ and \vaecsig. 
Leaving the noise as an additional optimization parameter decreases the performance from low \ac{snr} to approximately 10\,dB. 
From this point on, there is almost no gap between \vaee\ and \vaeesig, and \vaec\ and \vaecsig, highlighting that \vaeesig and \vaecsig estimate the noise variance with low error.

Fig.~\ref{fig:1-obs} presents the same evaluation as in Fig.~\ref{fig:5-obs} but for evaluation data with \textit{one} snapshot. 
We do not plot the results of the \ac{ic}-based methods here as they would work with a rank one sample covariance matrix.
In this case, the sample covariance matrix only has one non-zero eigenvalue, making the application of the \ac{ic}-based methods useless.
In Fig.~\ref{fig:1-obs}, the \acp{vae} that know about the noise variance can beat CovNet for every \ac{snr} value. 
Only \vaecsig\ performs worse than CovNet from 0 to 10\,dB. 
This shows that the proposed approach can also beat a supervised method in \ac{mo} selection if only a single snapshot is considered. 
Additionally, the performance gap between \vaee\ and \vaec\ is almost negligible for a single snapshot.

With \vaec\ and \vaecsig, we have low complexity methods that only require a forward pass through the neural networks and evaluation of the entropy routine from Section~\ref{subsec:mo-vae}. 
If it is possible to invest more computational complexity, \vaee\ and \vaeesig\ provide performance gains compared to their pendants \vaec\ and \vaecsig, respectively.

%% file: conclusion.tex
\section{Conclusion}
\label{sec:conclusion}

In this manuscript, we present methods for \ac{mo} selection based on a \ac{vae}. 
The main idea is to learn a parameterized covariance matrix at the \ac{vae} decoder. 
For the considered \ac{ula}, this is done with the help of an oversampled \ac{dft}-matrix. 
Simulation results highlight that the proposed methods are suitable for scenarios with very low snapshots, where \ac{ic}-based approaches fail. 
Comparisons with a supervised method highlight that the proposed methods can compete with a supervised approach and even beat it in a single snapshot scenario.
Interesting future steps include the investigation of other array geometries, correlated sources, and the optimal oversampling factor of the \ac{dft}-matrix.

%% file: main.bbl
\begin{thebibliography}{10}
\providecommand{\url}[1]{#1}
\csname url@samestyle\endcsname
\providecommand{\newblock}{\relax}
\providecommand{\bibinfo}[2]{#2}
\providecommand{\BIBentrySTDinterwordspacing}{\spaceskip=0pt\relax}
\providecommand{\BIBentryALTinterwordstretchfactor}{4}
\providecommand{\BIBentryALTinterwordspacing}{\spaceskip=\fontdimen2\font plus
\BIBentryALTinterwordstretchfactor\fontdimen3\font minus
  \fontdimen4\font\relax}
\providecommand{\BIBforeignlanguage}[2]{{%
\expandafter\ifx\csname l@#1\endcsname\relax
\typeout{** WARNING: IEEEtran.bst: No hyphenation pattern has been}%
\typeout{** loaded for the language `#1'. Using the pattern for}%
\typeout{** the default language instead.}%
\else
\language=\csname l@#1\endcsname
\fi
#2}}
\providecommand{\BIBdecl}{\relax}
\BIBdecl

\bibitem{Krim1996}
H.~Krim and M.~Viberg, ``{Two Decades of Array Signal Processing Research: The
  Parametric Approach},'' \emph{IEEE Signal Process. Mag.}, vol.~13, no.~4, pp.
  67--94, 1996.

\bibitem{Barthelme2021}
A.~Barthelme and W.~Utschick, ``{ChainNet: Neural Network-Based Successive
  Spectral Analysis},'' \emph{arXiv preprint arXiv:2105.03742}, 2021.

\bibitem{Stoica2004}
P.~Stoica and Y.~Selen, ``{Model-Order Selection: A review of information
  criterion rules},'' \emph{IEEE Signal Process. Mag.}, vol.~21, no.~4, pp.
  36--47, 2004.

\bibitem{Ding2018}
J.~Ding, V.~Tarokh, and Y.~Yang, ``{Model Selection Techniques: An Overview},''
  \emph{IEEE Signal Process. Mag.}, vol.~35, no.~6, pp. 16--34, 2018.

\bibitem{Wax1985}
M.~Wax and T.~Kailath, ``{Detection of Signals by Information Theoretic
  Criteria},'' \emph{IEEE Trans. Acoust.}, vol.~33, no.~2, pp. 387--392, 1985.

\bibitem{Bialer2019}
O.~Bialer, N.~Garnett, and T.~Tirer, ``{Performance Advantages of Deep Neural
  Networks for Angle of Arrival Estimation},'' in \emph{2019 IEEE Int. Conf.
  Acoust. Speech Signal Process.}\hskip 1em plus 0.5em minus 0.4em\relax
  Brighton, UK: IEEE, 2019, pp. 3907--3911.

\bibitem{Barthelme2020}
A.~Barthelme, R.~Wiesmayr, and W.~Utschick, ``{Model Order Selection in DoA
  Scenarios via Cross-entropy Based Machine Learning Techniques},'' in
  \emph{2020 IEEE Int. Conf. Acoust. Speech Signal Process.}\hskip 1em plus
  0.5em minus 0.4em\relax Barcelona, Spain: IEEE, 2020, pp. 4622--4626.

\bibitem{Barthelme2021a}
A.~Barthelme and W.~Utschick, ``{A Machine Learning Approach to DoA Estimation
  and Model Order Selection for Antenna Arrays with Subarray Sampling},''
  \emph{IEEE Trans. Signal Process.}, vol.~69, pp. 3075--3087, 2021.

\bibitem{Zhou2022}
S.~Zhou, T.~Li, Y.~Li, R.~Zhang, and Y.~Ruan, ``{Source Number Estimation via
  Machine Learning Based on Eigenvalue Preprocessing},'' \emph{IEEE Commun.
  Lett.}, vol.~26, no.~10, pp. 2360--2364, 2022.

\bibitem{Kingma2014}
D.~P. Kingma and M.~Welling, ``{Auto-Encoding Variational Bayes},'' in
  \emph{Proc. 2nd Int. Conf. Learn. Represent.}, Banff, Canada, 2014.

\bibitem{Xia2022a}
W.~Xia, S.~Rangan, M.~Mezzavilla, A.~Lozano, G.~Geraci, V.~Semkin, and
  G.~Loianno, ``{Generative Neural Network Channel Modeling for Millimeter-Wave
  UAV Communication},'' \emph{IEEE Trans. Wirel. Commun.}, vol.~21, no.~11, pp.
  9417--9431, 2022.

\bibitem{Caciularu2018}
A.~Caciularu and D.~Burshtein, ``{Blind Channel Equalization Using Variational
  Autoencoders},'' in \emph{2018 IEEE Int. Conf. Commun. Work.}\hskip 1em plus
  0.5em minus 0.4em\relax Kansas City, USA: IEEE, 2018, pp. 1--6.

\bibitem{Caciularu2020}
------, ``{Unsupervised Linear and Nonlinear Channel Equalization and Decoding
  Using Variational Autoencoders},'' \emph{IEEE Trans. Cogn. Commun. Netw.},
  vol.~6, no.~3, pp. 1003--1018, 2020.

\bibitem{Lauinger2022a}
V.~Lauinger, F.~Buchali, and L.~Schmalen, ``{Blind Equalization and Channel
  Estimation in Coherent Optical Communications Using Variational
  Autoencoders},'' \emph{IEEE J. Sel. Areas Commun.}, vol.~40, no.~9, pp.
  2529--2539, 2022.

\bibitem{Baur2022}
M.~Baur, B.~Fesl, M.~Koller, and W.~Utschick, ``{Variational Autoencoder
  Leveraged MMSE Channel Estimation},'' in \emph{2022 56th Asilomar Conf.
  Signals, Syst. Comput.}\hskip 1em plus 0.5em minus 0.4em\relax Pacific Grove,
  USA: IEEE, 2022, pp. 527--532.

\bibitem{Kingma2019}
D.~P. Kingma and M.~Welling, ``{An Introduction to Variational Autoencoders},''
  \emph{Found. Trends{\textregistered} Mach. Learn.}, vol.~12, no.~4, pp.
  307--392, 2019.

\bibitem{Liaw2018}
R.~Liaw, E.~Liang, R.~Nishihara, P.~Moritz, J.~E. Gonzalez, and I.~Stoica,
  ``{Tune: A Research Platform for Distributed Model Selection and Training},''
  \emph{arXiv preprint arXiv:1807.05118}, 2018.

\bibitem{Kingma2015}
D.~P. Kingma and J.~Ba, ``{Adam: A Method for Stochastic Optimization},'' in
  \emph{Proc. 3rd Int. Conf. Learn. Represent.}, San Diego, USA, 2015.

\end{thebibliography}
